\documentstyle[preprint,aps,tighten]{revtex}
\begin{document}
\draft

\title{Diffusive conductors as Andreev interferometers}

\author{ Yuli V. Nazarov and T. H. Stoof}
\address{
Faculteit der Technische Natuurkunde, Technische Universiteit Delft,
2628 CJ Delft, the Netherlands}
\maketitle
\begin{abstract}
We present a novel mechanism of phase-dependent
electric transport in diffusive normal metal-superconductor structures.
We provide a detailed theoretical and numerical analysis of recent
unexplained experiments essentially explaining them.
\end{abstract} 


What is the resistance of a small normal structure adjacent
to a superconductor? Superconductivity penetrates the structure
provided it is short enough. A naive suggestion would be
that the resistance vanishes. However, it is not so. The
simplest way to see this is to relate the resistivity to
the scattering in the structure.\cite{Scattering}
Normal electrons traversing the structure should undergo scattering
even if their wave functions are distorted by superconductivity.

If the structure is connected to two superconducting terminals
having different phases, the resistance of the structure
will depend on the phase difference. This provides the physical
background for what is called Andreev interferometry. There is
an outburst of interest in the topic. Different types of Andreev
interferometers have been proposed theoretically
\cite{Interferometers_theory,Circuit} and realized experimentally.
\cite{Pothier,Vegvar,Petrashov}

Andreev scattering reveals a significant difference between
diffusive conductors and tunnel junctions of the same resistance.
Optimal interferometers are composed of tunnel junctions.
\cite{Circuit,Pothier} Standard theory predicts that the zero-voltage,
zero-temperature resistance of a diffusive conductor is not
affected by penetrating superconductivity \cite{Noeffect},
although it is slightly modified at finite temperature, when
the sample length becomes comparable with the superconducting
correlation length in the normal metal, $\xi=\sqrt{D/(\pi T)}$,
$D$ being diffusivity. Apart from this thermal effect, a
small modification may arise from the weak localization correction.
\cite{Weak_localization}

However, the recent experiment \cite{Petrashov} demonstrates
a significant phase modulation of the resistivity in the diffusive
regime at very low temperatures. The authors insist that their
results can not be explained by existing theories.

Below we present a novel mechanism that provides zero-temperature
phase-dependent resistivity. Due to electron-electron interaction,
a weak pair potential is induced in the normal metal, that leads
to extra Andreev reflection occurring in the structure rather
than in the superconductors.
The relative change of the resistivity $\delta R/R \simeq g$, $g$ being
the interaction
parameter, and may be of either sign depending on the sign of $g$.

However, our numerical results shows that for the concrete structure
its magnitude is too small to explain the data of Ref. \cite{Petrashov}.
Careful analysis allowed us to conclude that actually the more
trivial thermal effect has been observed. The data show
excellent agreement with the results of our simulations.

The most adequate theoretical description of the system
is provided in the framework of the Keldysh
Green's function technique elaborated in \cite{Larkin} for
superconductivity. In the diffusive approximation, one  first gets
equations for the advanced (retarded) Green's function, which
is a $2 \times 2$ matrix depending on coordinate and energy,
$\hat G(x,\epsilon)$, with $\hat G^2 = \hat 1$,
\begin{equation}
\partial_x(D\hat G \partial_x) +i [\hat H,\hat G]=0.
\label{advanced}
\end{equation}
$D$ being the diffusivity in the normal state, $\hat H = \epsilon \sigma_z
+ i ({\rm Re} \Delta(x) \sigma_y + {\rm Im} \Delta(x) \sigma_x)$.We will assume
that the temperature is low enough and the size of the
normal structure is large enough to satisfy $T,D/L^2 \ll \Delta_S$,
$\Delta_S$ being the energy gap in the superconductors.  Then the boundary conditions
for $\hat G$ take a simple form ($\sigma$ being Pauli matrices):
$\hat G = \sigma_z$ in normal reservoirs, and $\hat G=\sigma_x\sin \phi +
\sigma_y \cos \phi $ in a superconducting reservoir having phase $\phi$.

The Green's function $\hat G$ determines the characteristics of
the energy spectrum of the quasiparticles in the structure. To solve
the transport problem, we must know how this spectrum is filled by
extra quasiparticles.
The equation for the even-in-energy part of the filling factor reads
\begin{equation}
 \partial_x(D(\epsilon,x) \partial_x f(\epsilon,x)) +\gamma(x)f(\epsilon)=0.
\label{filling}
\end{equation}
The first term here describes diffusion of quasiparticles with a diffusion
coefficient that is changed by the penetrating superconductivity, $D(\epsilon)=
D {\rm Tr}((\hat G +\hat G^+)^2)/8$. At zero temperature, $\hat G=\hat G^+$
and $D$ remains unchanged. The second term describes absorption of
quasiparticle excitations into the superconducting condensate, or,
in other words, conversion of the normal current into the superconducting
one.
The coefficient $\gamma$ is proportional to the local value of
the pair potential, $\gamma(x)=\Delta(x)
\rm{Tr}[i \sigma_y (\hat G(\epsilon,x)+\hat G(-\epsilon,x))] $.

In a normal reservoir biased at voltage $V$ with respect to superconductors
$f(\epsilon)=eV/4T \cosh^{-2}(\epsilon/2T)$. This provides boundary
conditions for (\ref{filling}). The current into a reservoir is determined by
the local gradient of $f$.

The common theoretical approach
(see, for instance, \cite{Klapwijk,Scattering,Circuit})
disregards interactions in the normal metal, that leads
to $\Delta,\gamma \equiv 0$. Since for diffusive conductors at zero
temperature the common non-interacting picture does not
give the resistance change, we concentrate on the effect
of $\Delta$ in the normal metal. This value can be calculated
with
\begin{equation}
\Delta= g \int d \epsilon  \tanh(\epsilon/2T) {\rm Tr} [i \sigma_y
(\hat G^A - \hat G^R)]/8.
\label{delta}
\end{equation}
This is the novel feature of the present approach.

Let us first make a simple estimation of the magnitude of the
effect. Since it is expected to be small, we solve
Eq. \ref{filling} to first order in $\gamma$. This gives a relative
resistance change $\delta R/R \simeq \gamma L^2/D$.
At zero energy, $\gamma \simeq \Delta$. In normal metal,
the energies in the window $\simeq D/L^2$ contribute to $\Delta$,
therefore $\Delta \simeq g D/L^2$. This results in a simple
estimation for the resistance change,
\begin{equation}
\delta R/R = g \ c(\Phi),
\label{estimation}
\end{equation}
$c$ being a dimensionless number depending on  the geometry
of the structure and on the distribution of the resistivity therein.
It is important to note that $c$ depends
neither on the structure size nor on the absolute value of
the resistivity, provided the temperature is low enough, $T \ll D/L^2$.
The effect depends on the normal metal material by means of $g$
and can be of either sign depending on the sign of $g$.
If the geometry of the structure is well defined,
the effect can be used for the direct measurement of interaction
in normal metal.

At qualitative level, the effect seems to explain the results of
Ref. \cite{Petrashov}. Indeed, the phase modulation of the
resistance they observed at low temperatures
was of the order of several percent
and looked material-dependent, including the sign of the resistance
change. This prompted us to detailed numerical calculations of
the resistance of a concrete structure ( Fig. 1) which is
very similar to the one used in Ref. \cite{Petrashov}.

The structure consists of the current branch, the superconducting
branch connected to superconducting reservoirs biased
at the phases $-\Phi/2$,$\Phi/2$ respectively, and the extra
branch made for technological reasons. The current flows
as it is shown in Fig.2 and the voltage difference between
the points $A$ and $A'$ is  measured. Owing to the symmetry
of the structure, the voltage and $f$ distributions are antisymmetric
with respect to the superconducting branch, whereas $\Delta(x)$ and $\hat G(x)$
are symmetric. Superconductivity in the structure gets
completely suppressed when $\Phi$ approaches $\pi$.

One-dimensional differential equations (\ref{advanced}), (\ref{filling}) shall be solved
for each branch and then matched in the crossing points.
Fist we calculate $\hat G(\epsilon) $ in all points of the structure.
Due to the boundary conditions, it depends on $\Phi$ in each point,
thus providing the origin of the phase dependent effect.
We obtain $\Delta(x)$ by integrating $\hat G$ over energy.
Then we calculate $\gamma(x)$ and make use of an analytical
formula that relates the resistance change to $\gamma(x)$.
Details of the calculations will be reported elsewhere.
\cite{Elsewhere}

In Fig. 2 we have shown the calculated phase dependence normalized
by its maximal value at $\delta\phi=0$, $c(\phi)/c(0)$, contrasted
with the experimental data for Ag. The phase dependencies look
similar, but the magnitude of the effect cannot be satisfactory
explained. According to the calculation, $c(0)=0.14$. If we
take an expected value $g=0.04$ \cite{Mota} for silver, we would obtain
$\delta R_{max}/R=0.003$ whereas the experiment gives
$\delta R_{max}/R=0.1$.
If we do it the other way around and try to fit $g$ from the experiment
we end up with $g \simeq 0.7 $. That would bring silver to the
rank of high temperature superconductors.

This prompted us to check the possibility that had been rejected
by the authors of Ref. \cite{Petrashov}. We have calculated the
thermal effect on the resistivity neglecting interaction corrections.
For a given geometry, the relative resistance change $\delta R/R$
is a function of $L/\xi$ and $\phi$. Our results are presented in Fig. 3.
As  expected, the thermal effect vanishes both at low and high
temperatures. The resistance at $\Phi=0$ reaches the minimum
at $ L \approx 3 \xi$. We have plotted the normalized phase dependence
at $L=3\xi$ in Fig. 4 along with experimental data and obtained
a perfect match. The maximal values of the change are also
very close to each other: $(\delta R_{max}/R)_{theor}=-0.097$
versus  $(\delta R_{max}/R)_{exp}=-0.11$. From the estimations given
in \cite{Petrashov} we obtain indeed $L/\xi=2.5-3$ at $T=20 mK$.
As we can see in Fig. 3, the thermal effect persists at rather
high temperature, in agreement with the long high
temperature tail observed in \cite{Petrashov}.

All this allows us to conclude that the experiment \cite{Petrashov}
can be perfectly described by existing semiclassical theory of
superconducting proximity effect and thus to remedy the
seeming discrepancy between theory and experiment.

The remaining discrepancy for metallic samples
can be easily understood if one  take
into account the sensitivity of the effect to a concrete
geometry and to the inevitable inhomogeneity
of these ultra-small structures. This point of view
is supported by large sample-to-sample fluctuations
of the magnitude of the effect. \cite{Petrashov}
The understanding of the results for $Sb$ samples
having high resistivity presents a certain difficulty.
However, we notice that all essential features
for these samples, such a small magnitude of the
effect, sinusoidal phase dependence, positive sign,
and the long high-temperature tail, can be well understood
if the structures are not completely diffusive but
contain tunnel junctions. \cite{Circuit}

We are ready to present several conclusions.

We have shown that the results of \cite{Petrashov}
can be perfectly explained within the existing
theoretical framework and be attributed to the thermal
effect, provided a concrete experimental geometry
is taken into account.

We present a novel mechanism  of phase-dependent
resistance in hybrid normal metal-superconductor
structures that works at zero temperature.
The observation of this effect would allow a
direct measurement of the interaction parameter
in a normal metal.

Our results show that the observation of
the weak localization  correction \cite{Weak_localization}
is a more difficult task than it was thought to be.
At zero temperature, this correction will be masked
by the novel effect we discussed, provided the interaction
is not very small. At high temperatures,
the correction would become comparable with the tail of
the thermal effect at $T > D R_Q/L^2 R$. For the structures
used in Ref. \cite{Petrashov} this would correspond to
unreasonably high temperatures of $20 {\rm K}$.

The authors are indebted to V. Petrashov for numerous discussions
of his results, B. Z. Spivak for his illuminating remarks concerning
the interaction in normal metal, D. Esteve, M. Devoret, H. Pothier and S. Gueron
for the discussion from which this work has emerged, A. V.
Zaitsev and  G. E. W. Bauer for helpful comments.
This work is a part of the research program of the "Stichting voor
Fundamenteel Onderzoek der Materie"~(FOM), and we acknowledge the financial
support from the "Nederlandse Organisatie voor Wetenschappelijk Onderzoek"
~(NWO).

\begin{figure}
\caption{The structure under consideration.}
\label{fig1}
\end{figure}

\begin{figure}
\caption{
Normalized phase dependence of the novel effect.
Squares correspond to the experimental data of Ref. ~6.
}
\label{fig2}
\end{figure}

\begin{figure}
\caption{Temperature and phase dependence of the thermal effect.
The temperature is incorporated into $\xi^2 = D/\pi T$.
The phase difference changes from $0$ for the lowermost
curve to $\pi$ for the uppermost one with  step $\pi/20$.}
\label{fig3}
\end{figure}

\begin{figure}
\caption{
Normalized phase dependence of the thermal effect
at $L=3\xi$.
Squares correspond to the experimental data of Ref. ~6.
}
\label{fig4}
\end{figure}


\begin{references}
\bibitem{Scattering}
C. J. Lambert, J. Phys.:Condens. Matter, {\bf 3}, 6579 (1991);
C. W. J. Beenakker, Phys. Rev. B {\bf 46}, 12841 (1992).
\bibitem{Interferometers_theory}
H. Nakano and H. Takayanagi, Sol. St. Comm. {\bf 80}, 997 (1991);
F. W. J. Hekking and Yu. V. Nazarov, Phys. Rev. Lett. {\bf 71}, 1625 (1993);
A. V. Zaitsev, Phys. Lett. A {\bf 194}, 315 (1994);
A. Kadigrobov, A. Zagoskin, R. I. Shekhter, and M. Jonson, preprint, 1995;
N. K. Allsopp, J. Sanchez Canizares, R. Raimondi, and C. J. Lambert,
preprint, 1995.
\bibitem{Circuit} Yu. V. Nazarov, Phys. Rev. Lett. {\bf 73}, 1420 (1994).
\bibitem{Pothier} H. Pothier, S. Gueron, D. Esteve, Phys. Rev. Lett.
{\bf 73}, 2488 (1994).
\bibitem{Vegvar} P. G. N. de Vegvar, T. A. Fulton, W. H. Mallison,
and R. E. Miller, Phys. Rev. Lett. {\bf 73}, 1416 (1994);
A. Dimoulas, J. P. Heida, B. J. van Wees, T. M. Klapwijk,
W. v. d. Graaf, and G. Borghs, Phys. Rev. Lett. {\bf 74}, 602 (1995).
\bibitem{Petrashov} V. T. Petrashov, V. N. Antonov, P. Delsing and
T. Claeson, Phys. Rev. Lett. {\bf 74}, 5268 (1995).
\bibitem{Noeffect} S. N. Artemenko, A. F. Volkov, and A. V. Zaitsev,
Sol. St. Comm. {\bf 30}, 771 (1979);
see also Refs. \cite{Scattering,Circuit}.
\bibitem{Weak_localization} B. Z. Spivak and D. E. Khmelnitskii,
JETP Lett. {\bf 35}, 412 (1982)
[Pis'ma Zh. Eksp. Teor. Fiz. {\bf 35}, 334 (1982) ].
\bibitem{Larkin} A. I. Larkin and Yu. V. Ovchinninkov,
 Zh. Eksp. Teor. Fiz. {\bf 68}, 1915 (1975)
[Sov. Phys. JETP {\bf 41}, 960 (1975)];
 A. I. Larkin and Yu. V. Ovchinninkov,
 Zh. Eksp. Teor. Fiz. {\bf 73}, 299 (1977)
[Sov. Phys. JETP {\bf 46}, 155 (1977)];
\bibitem{Klapwijk} A. F. Volkov, A. V. Zaitsev, and T. M. Klapwijk,
Physica C {\bf 210}, 21 (1993).
\bibitem{Mota} A. C. Mota, P. Visani, and A. Pollini,
Jorn. of Low Temp. Phys., {\bf 76}, 465 (1989).
\bibitem{Elsewhere} T. H. Stoof and Yu. V. Nazarov,
manuscript in preparation.
\end{references}
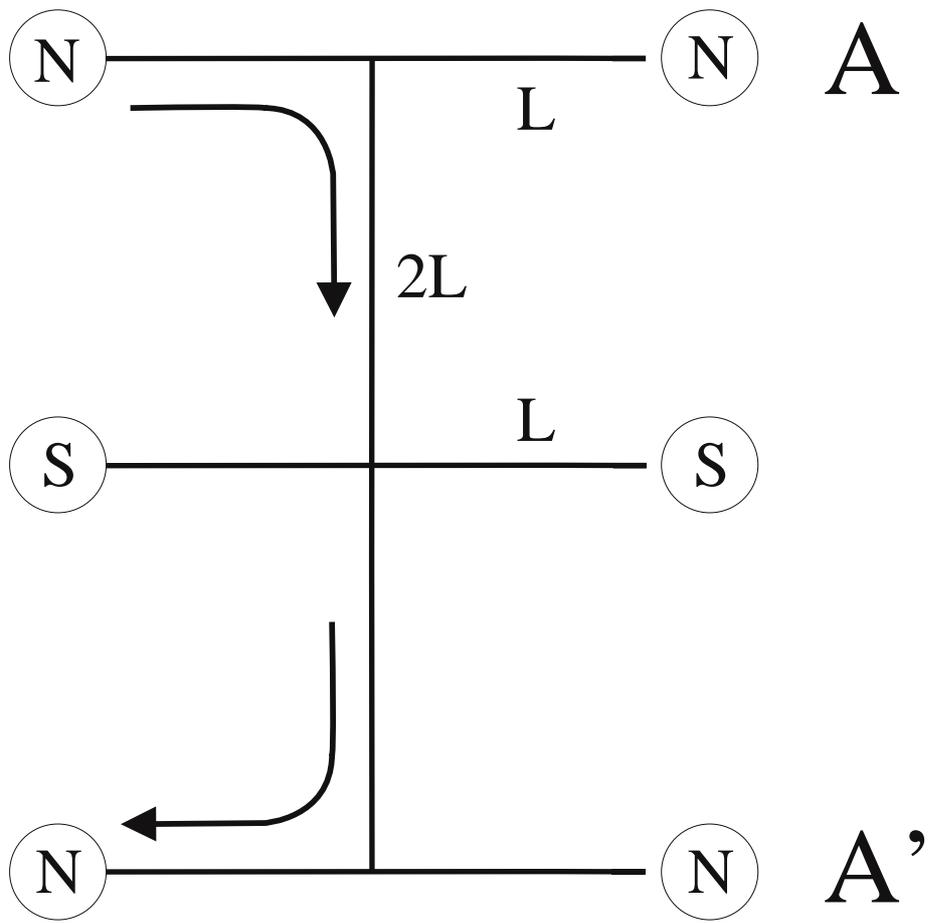
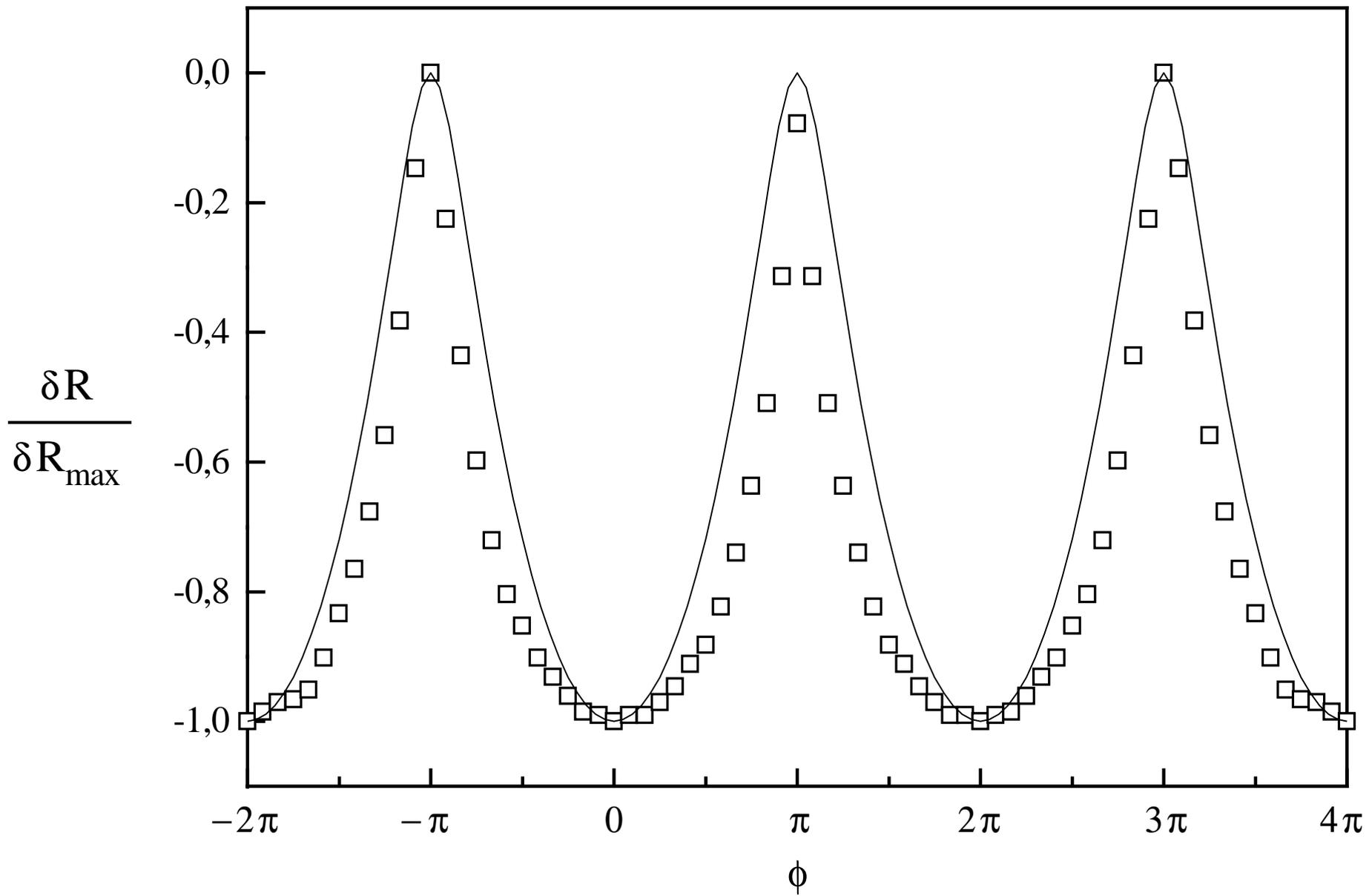
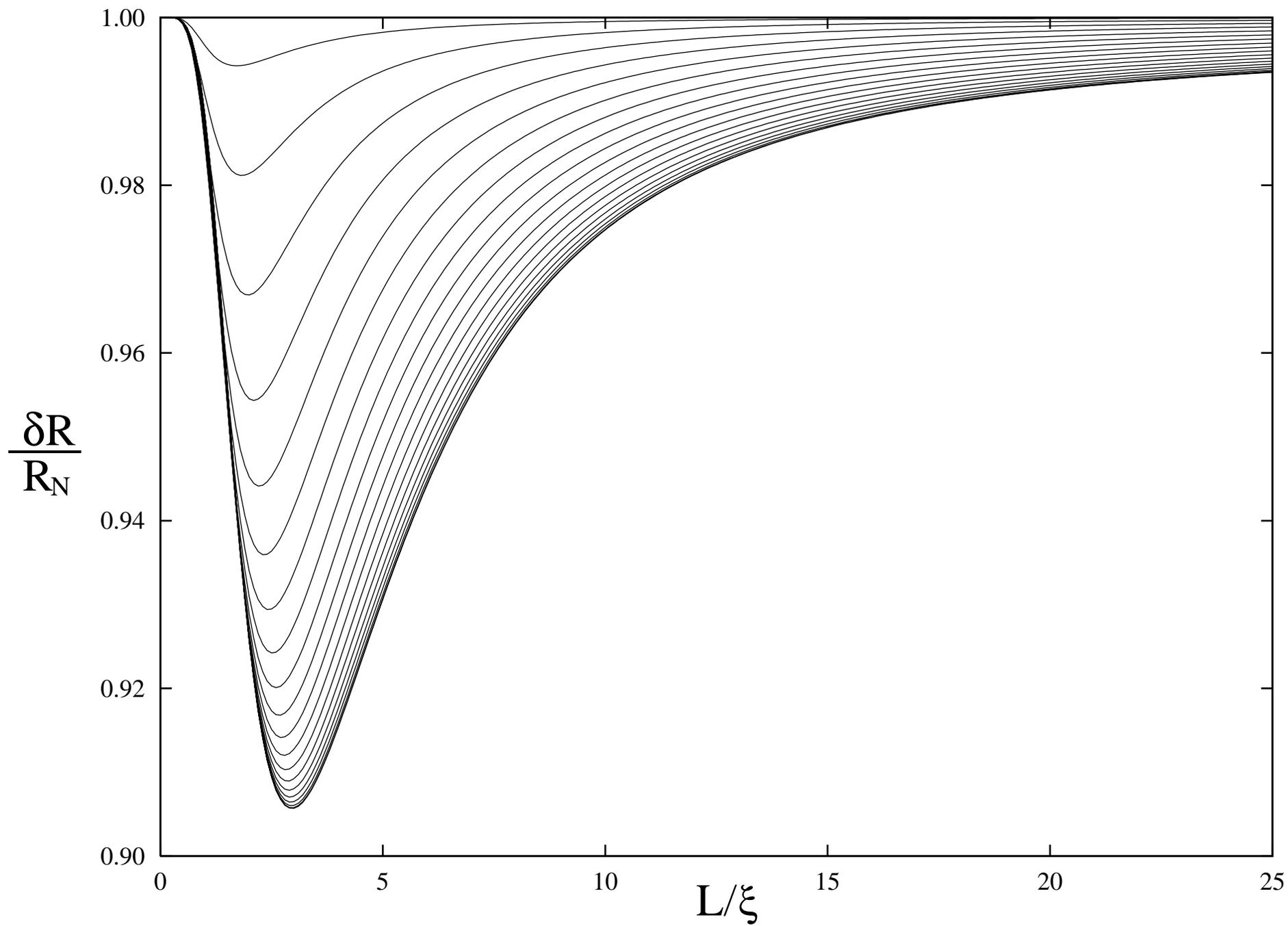
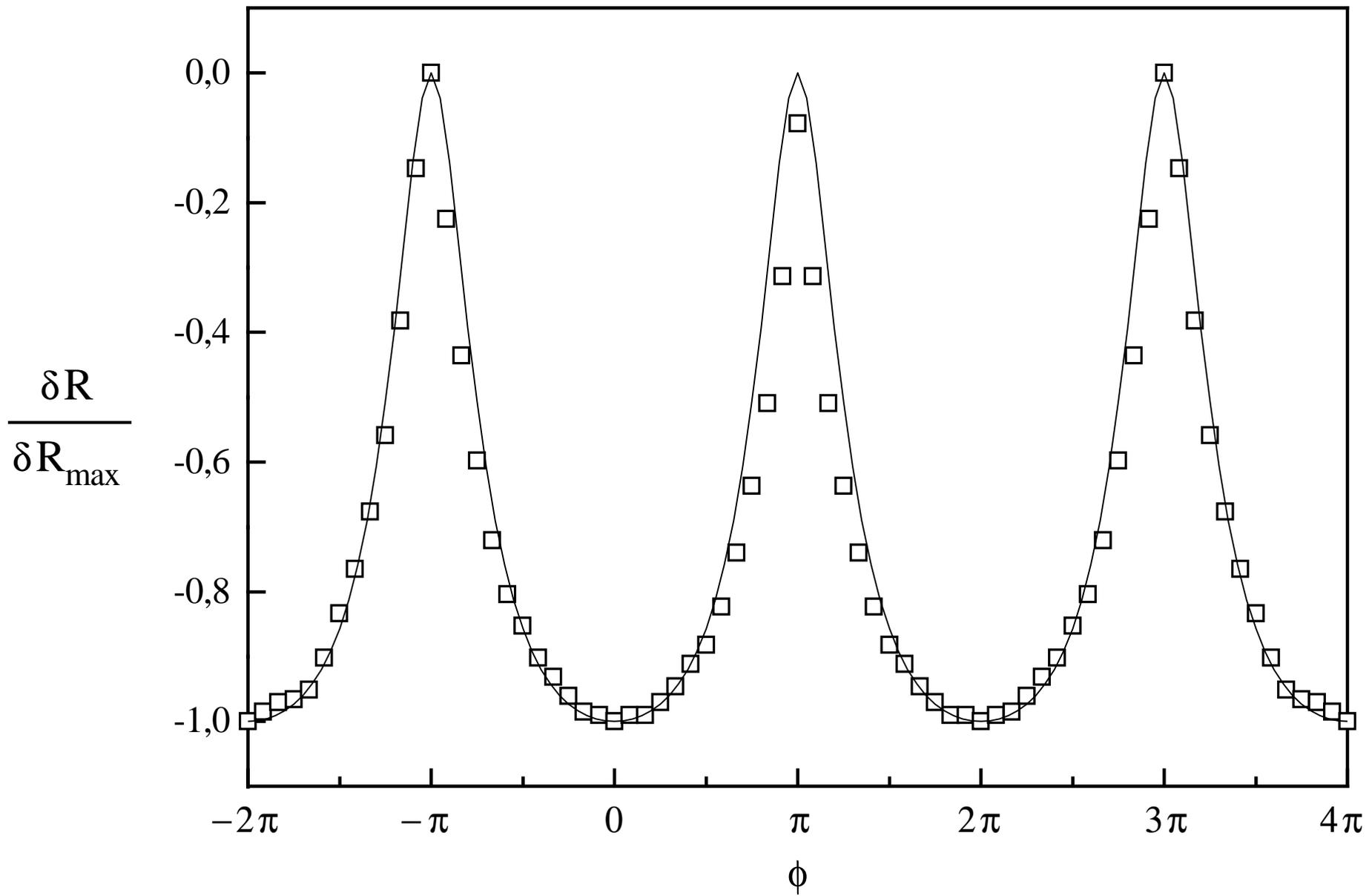
\end{document}